\documentclass{aa}  

\usepackage{graphicx}
\usepackage{txfonts}
\usepackage{color}

\begin{document} 

   \title{Deconstructing the Antlia cluster core}

   \titlerunning{Deconstructing the Antlia cluster core}

   \author{J. P. Caso
          \inst{1,2}
          \and
          T. Richtler
          \inst{3}
           }

\authorrunning{Caso \& Richtler}

   \institute{Grupo de Investigaci\'on CGGE, Facultad de Ciencias Astron\'omicas y Geof\'isicas de la Universidad Nacional de La Plata,    
and \\ Instituto de Astrof\'isica de La Plata (CCT La Plata -- CONICET, UNLP), Paseo del Bosque S/N,  
B1900FWA La Plata, Argentina
       \and
Consejo Nacional de Investigaciones Cient\'ificas y T\'ecnicas, Rivadavia 1917, C1033AAJ  
Ciudad Aut\'onoma de Buenos Aires, Argentina
         \and
Departamento de Astronom\'{\i}a,
Universidad de Concepci\'on,
Concepci\'on, Chile}

   \date{Received ...; accepted ...}

 
  \abstract
   {The present literature does not give a satisfactory answer to the question about
   the nature of the "Antlia galaxy cluster".}
   {The radial velocities of galaxies found in the region around the giant ellipticals NGC 3258/3268 range from about 1000 km/s to 4000 km/s.
    We characterise this region  and its possible kinematical and population substructure.
}
   {We have obtained VLT--VIMOS multi-object spectra of the galaxy
population in the inner part of the Antlia cluster and measure radial velocities for 45 potential members. We supplement 
our galaxy sample with literature data, ending up with 105 galaxy 
velocities.}
   {We find a large radial velocity 
dispersion for the entire sample as reported in previous papers. 
However, we find three groups at about 1900 km/s, 2800 km/s, and 3700 km/s, which
we interpret as differences in the recession velocities rather than peculiar 
velocities.
 }
 { The high radial velocity dispersion of galaxies in the Antlia region reflects a
 considerable extension along the line of sight. 
}

   \maketitle
%

\section{Introduction}
\label{sec:intro}

The character of the galaxy assembly in the constellation of Antlia is 
not as clear as, for example, in the cases of the Virgo or Fornax galaxy 
clusters. Some authors used the term "galaxy group" \citep{fer90}, others 
"galaxy cluster" \citep{hop85,smi08a}, and
\citet{hop85} identified five "clusters" in the Antlia region ($\alpha= 
10^h - 10^h\,50^m$, $\delta= -42^o - -30^o$). The most striking central 
cluster is called "Antlia II". Its morphological appearance is 
that of two groups concentrated around the giant ellipticals NGC\,3258 
and NGC\,3268 in a projected distance of 220\,kpc.

Both dominant galaxies 
are extended X-ray sources \citep{nak00,ped97}
 that exhibit rich globular cluster systems. 
The globular cluster system of NGC\, 3258 contains about 6000 members.
The system of 
NGC\,3268 is somewhat poorer, but is still typical of a giant elliptical 
galaxy, with almost 5000 globular clusters \citep{bas08}.
\citet{hop85} and, more recently, \citet{hes15} estimated velocity 
dispersions of $\sim 500$\,km\,s$^{-1}$ for the bright population 
of Antlia,  higher than those in clusters like Fornax \citep{dri01}.
The  globular cluster luminosity function indicated that NGC\,3258 
could be a few Mpc nearer than NGC\,3268 \citep{dir03a,bas08}, which agrees
with the distances obtained with surface brightness fluctuations by 
\citet{bla01}. However, \citet{can05} quoted the same distance
moduli for both galaxies.

Despite being nearby, a thorough radial velocity survey of Antlia's 
galaxy population has not been done yet. In this paper, we present 
new radial velocities of galaxies located in
Antlia. We supplement our galaxy sample with literature data, ending 
up with 105 galaxy velocities. These data will help us to 
better understanding the structure of Antlia.

In the following, we keep the term "Antlia cluster" for 
simplicity. To maintain consistency with earlier papers 
\citep[e.g.][]{smi12,cas13a,cas14,cal15}, we adopt a distance 
of 35 Mpc, which means a scale of 169.7 pc/arcsec.

\section{Observations and reductions}
\label{sec:reduc}
\begin{figure}    
\includegraphics[width=90mm]{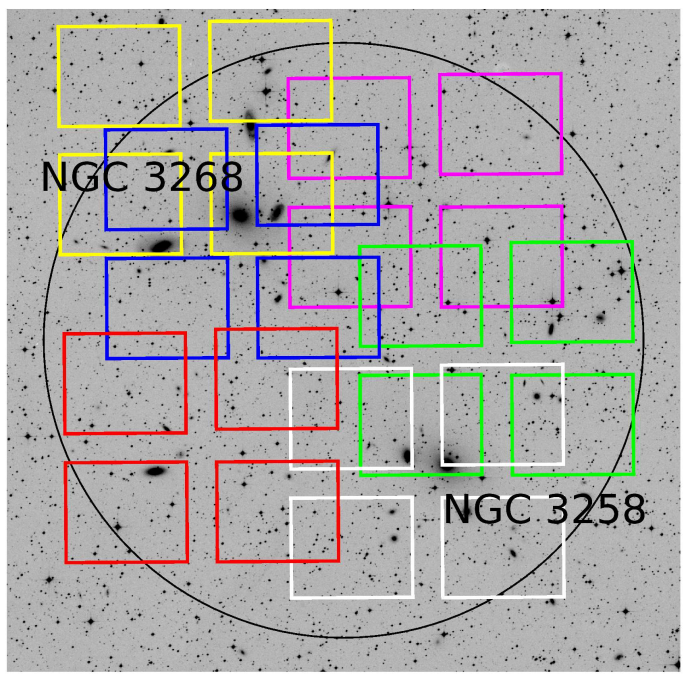}    
\caption{Positions of the six VIMOS fields (each one composed of four 
quadrants). The black circle is centred on the midpoint between the 
projected positions for the two gEs, and its radius is $20'$. North is 
up, east to the left.}    
\label{campos}    
\end{figure}

We performed  multi-object spectroscopy with VLT-VIMOS  of the galaxy 
population in six fields located in the inner part of the Antlia cluster. 
The observations  were carried out under the  programmes 60.A-9050(A) and 
079.B-0480(B) (PI Tom Richtler), observed during the first semesters of 2007 
and 2008. 
Figure\,\ref{campos} shows the four quadrants for each one of the six VIMOS 
fields, using different colours.

The grating was HR blue, and the slit width was $1''$. 
For each science field, the integration time was 1~h, split into 
three individual exposures. This configuration implied a wavelength coverage 
spanning $3700\,\rm{\AA}  - 6600\,\rm{\AA}$ (depending on the slit positions) 
and a spectral resolution of $\sim 2.5\,\rm{\AA}$ .
The data were reduced with \textsc{esorex} in the usual manner for VIMOS 
data. First, a master bias was obtained for each field with the recipe VMBIAS  
from five individual bias exposures. The normalised master flat field was 
created with the recipe VMSPFLAT from a set of dome flat-field exposures. The 
recipe VMSPCALDISP was used to determine the wavelength calibrations and 
spectral distortions. Typically, more than 20 lines were 
identified for each slit. Afterwards, the bias and flat-field corrections were applied 
to each science exposure, together with the wavelength calibration. This 
was done with the recipe VMMOSOBSSTARE. Individual exposures were then 
combined with the \textsc{iraf} task IMCOMBINE to achieve a higher S/N. The 
spectra were extracted  with the task APALL, also within \textsc{iraf}.
We measured the heliocentric radial velocities using the \textsc{iraf} task 
FXCOR within the NOAO.RV package. We used synthetic templates, which were selected from 
the single stellar population (SSP) model spectra at the \textsc{miles} 
library (${\rm http://www.iac.es/proyecto/miles}$, \citealt{san06}).
We selected  SSP models with the metallicities [M/H]=\,-0.71 and 
[M/H]=\,-0.4, a unimodal initial mass function with slope $1.30$, and an age 
of 10\,Gyr. The wavelength coverage of these templates is 
$3700\,\rm{\AA} - 6500\,\rm{\AA}$, and their spectral resolution is 
$3\,\rm{\AA}$ FWHM.

We also obtained GEMINI-GMOS multi-object spectra from programme GS-2013A-Q-37
(PI J. P. Calder\'on).  The grating B600\_G5303 blazed at 
$5000\,\rm{\AA}$ was used  with a slit width of 1\,arcsec. The wavelength coverage
spans $3300\,\rm{\AA} - 7200\,\rm{\AA}$, depending on the position of the slits. The
data were reduced using the GEMINI.GMOS package within {\sc IRAF}. We refer
to \citet{cas14} for more information about the reduction. 

We could determine heliocentric radial velocities ($V_{R,h}$) for 67 
galaxies located in the inner $20'$ of the Antlia cluster (i.e.,
the inner 200\,kpc for our adopted distance). In previous 
studies \citep{smi08a,smi12,cas13a}, those galaxies with $V_{R,h}$  
 between 1200\,km\,s$^{-1}$ and 4200\,km\,s$^{-1}$ have been assigned 
 to Antlia (which already raised doubts owing to the large velocity 
interval). In our enhanced sample, we find the lowest velocity to be  
$V_{R,h} = 1150$\,km\,s$^{-1}$, and  there are no velocities between  
4300\,km\,s$^{-1}$ and $\sim7600$\,km\,s$^{-1}$. Galaxies with higher
$V_{R,h}$ than the latter limit were rejected from our sample and are
listed in Table\,\ref{tab.backg}. It can be noticed that several galaxies 
from the \citet{fer90} catalogue are indeed in the background. In these
cases, \citet{fer90} classified them as ``likely members'' or
``probable background''.
From the 67 galaxies with $V_{R,h}$ measurements, 45 are thus members of
our sample. 

\begin{figure}  
\includegraphics[width=90mm]{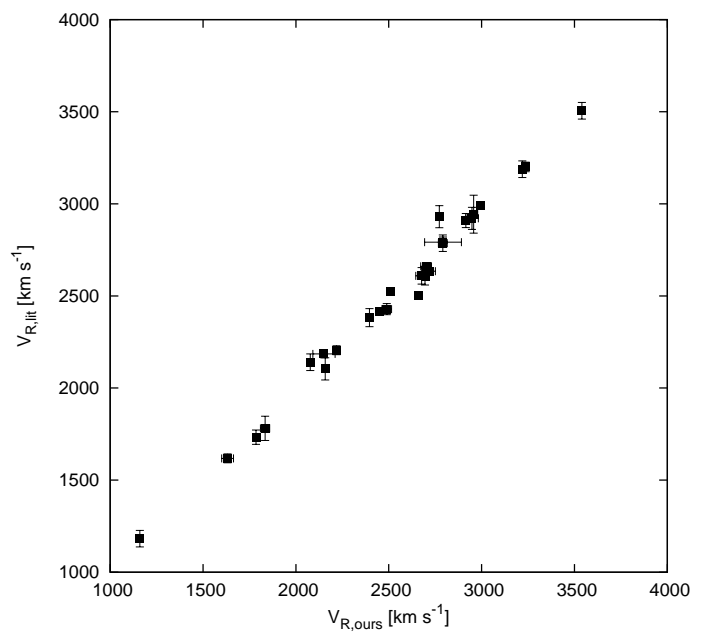}    
\caption{Comparison between heliocentric radial velocity measurements from 
this paper and the literature for Antlia members.}    
\label{vr.comp}    
\end{figure}

Additional $V_{R,h}$ measurements were collected from the
literature \citep[]{smi08a,smi12}
and the NED\footnote{This research has made use of the NASA/IPAC 
Extragalactic Database (NED), which is operated by the Jet Propulsion 
Laboratory, California Institute of Technology, under contract with the 
National Aeronautics and Space Administration.}.
 The final sample of Antlia members consists of 105 galaxies, which are 
listed in Table\,\ref{tab.mem}. A fraction of the present galaxies had
 been measured earlier. In these cases, we find good agreement between
our measurements and the literature values (Figure\,\ref{vr.comp}).
The mean $V_{R,h}$ difference and dispersion are 20\,km\,s$^{-1}$ and 
50\,km\,s$^{-1}$, respectively. In comparison, the uncertainties of our 
measurements are typically in the range $10-40$\,km\,s$^{-1}$.

\begin{figure}  
\includegraphics[width=90mm]{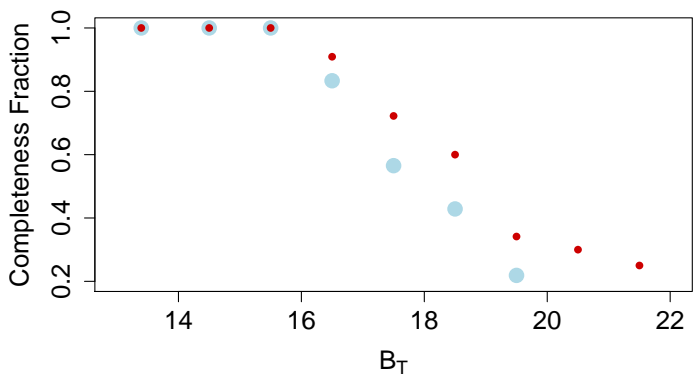}    
\caption{Completeness analysis for the spectroscopic sample considering
the entire photometric catalogue (large light-blue circles), and
excluding galaxies with \citet{fer90} membership status `3' (small red 
circles). The bin width is 1\,mag.}    
\label{compl}    
\end{figure}

\citet{fer90} provided the photometrical catalogue of galaxies in
Antlia with the largest spatial coverage, while \citet{cal15} have carried 
out a deeper survey of the early-type galaxies in a region that
contains our VIMOS fields. To supplement the \citet{fer90}
catalogue, we derived $B$ magnitudes for the galaxies measured
by \citet{cal15} in the Washington $(C,T_1)$ photometric system 
\citep{can76}. As a result, we applied equation\,4 from \citet{smi08a} to 
transform $(C-T_1)_0$ into $(B-R)_0$ colours. Then we obtained $B$ 
magnitudes, considering that $R$ and $T_1$ filters only differ in a 
small offset \citep{dir03a}. Figure\,\ref{compl} shows the completeness
for those galaxies located within $30'$ of the cluster
centre (see Section\,\ref{morph}), which roughly matches the region
observed with our VIMOS fields and previous spectroscopic studies 
\citep{smi08a,smi12}. The bin width is 1\,mag. When we exclude 
galaxies with \citet{fer90} membership status `3' (which are the less 
probable members), the $80\%$ completeness is reached 
at $B_T=17$\,mag and the $60\%$ at $B_T=19$\,mag (small red circles).

\section{Results}
\label{sec:resul}

\subsection{$V_{R,h}$ velocity distribution}
\label{sec:dvel}
\begin{figure}    
\includegraphics[width=90mm]{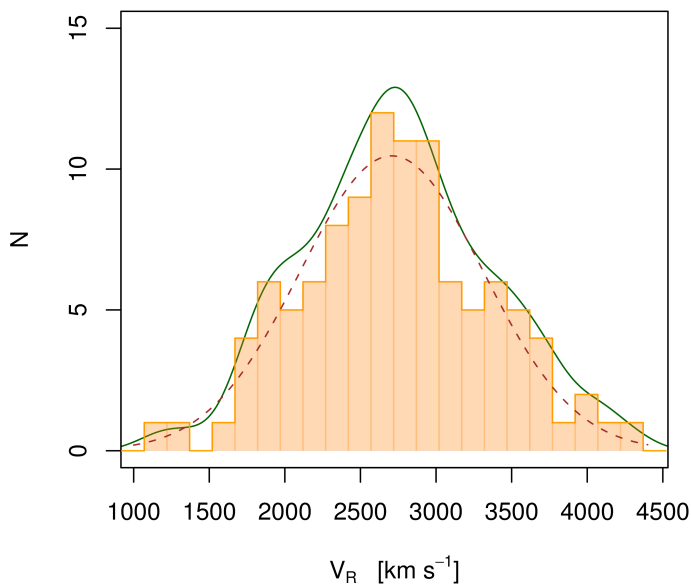}    
\caption{Histogram of the $V_{R,h}$ distribution for the galaxies in our Antlia 
sample. The bin width is 150\,km\,s$^{-1}$. Overplotted are the
smooth velocity distribution, obtained with a Gaussian kernel (green 
solid curve), and the normal profile fitted by least squares (brown 
dashed curve).}    
\label{dvel}    
\end{figure}

The histogram of the $V_{R,h}$ distribution for the galaxies in our Antlia 
sample is shown in Figure\,\ref{dvel} with a bin width of 150\,km\,s$^{-1}$. 
The green solid curve represents the smooth velocity distribution, obtained 
with a Gaussian kernel. 
The distribution resembles a Gaussian, which would be expected if 
the spatial velocity for the Antlia members is described by a Maxwellian 
distribution. 

A Shapiro-Wilk normality test returns a p value of 0.9
\citep[][hereafter S-W test]{sha65,roy95}, meaning that we cannot 
reject the hypothesis that our sample is drawn from a normal distribution.
Considering this, we fitted a normal profile to the $V_{R,h}$ distribution 
by least-squares, assuming Poisson uncertainties for the bin values. 
The best fit corresponds to a Gaussian distribution with a mean velocity 
of $2708\pm42$\,km\,s$^{-1}$ and a dispersion of $617\pm46$\,km\,s$^{-1}$ 
(dashed brown curve in Fig.\,\ref{dvel}), which agree with the 
values from optical data by \citet{hes15}. 
This dispersion fits massive clusters like Virgo \citep{con01}. 
\citet{nak00} obtained a gas temperature of 2\,keV for the surroundings 
of NGC\,3268, which does not exclude a dispersion of 617 km/s, considering 
the $\sigma$-T- relation of \citet{xue00} and its scatter.
However, whether this high velocity dispersion obtained for our Antlia 
sample represents a single virialized system, is an interesting
question, which we try to answer here.

Assuming that the individual $V_{R,h}$ uncertainties are the sigma values
of a normal distribution, we simulated the $V_{R,h}$ for each one of the 105 galaxies 
by a Monte-Carlo method. 
The procedure was repeated 100 times, in all cases obtaining a
non-rejection of the normality hypothesis. This demonstrates that the
intrinsic uncertainties of the measurements do not play a central
role in the result.

\begin{figure*}[ht!]
\includegraphics[width=60mm]{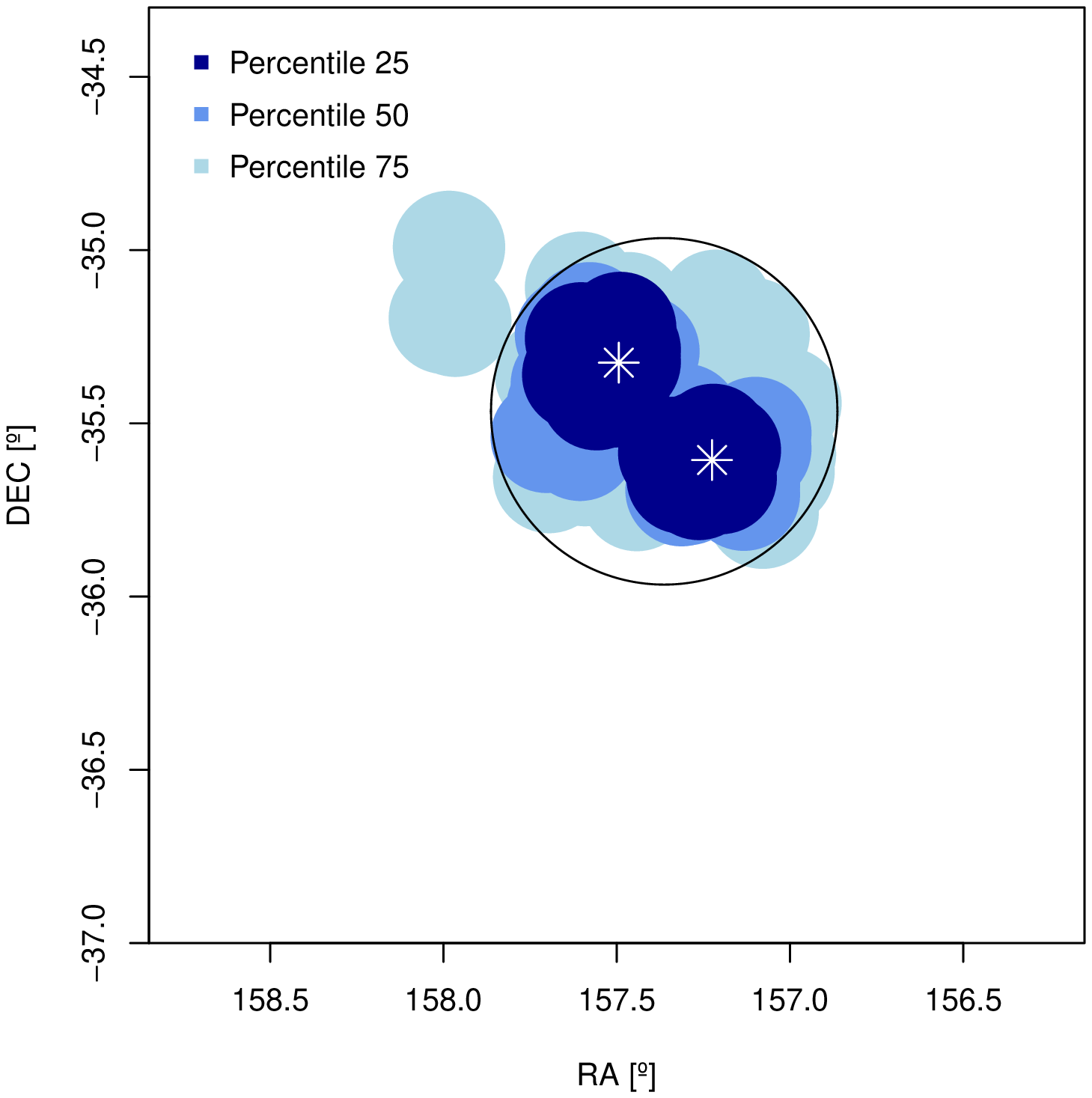}    
\includegraphics[width=60mm]{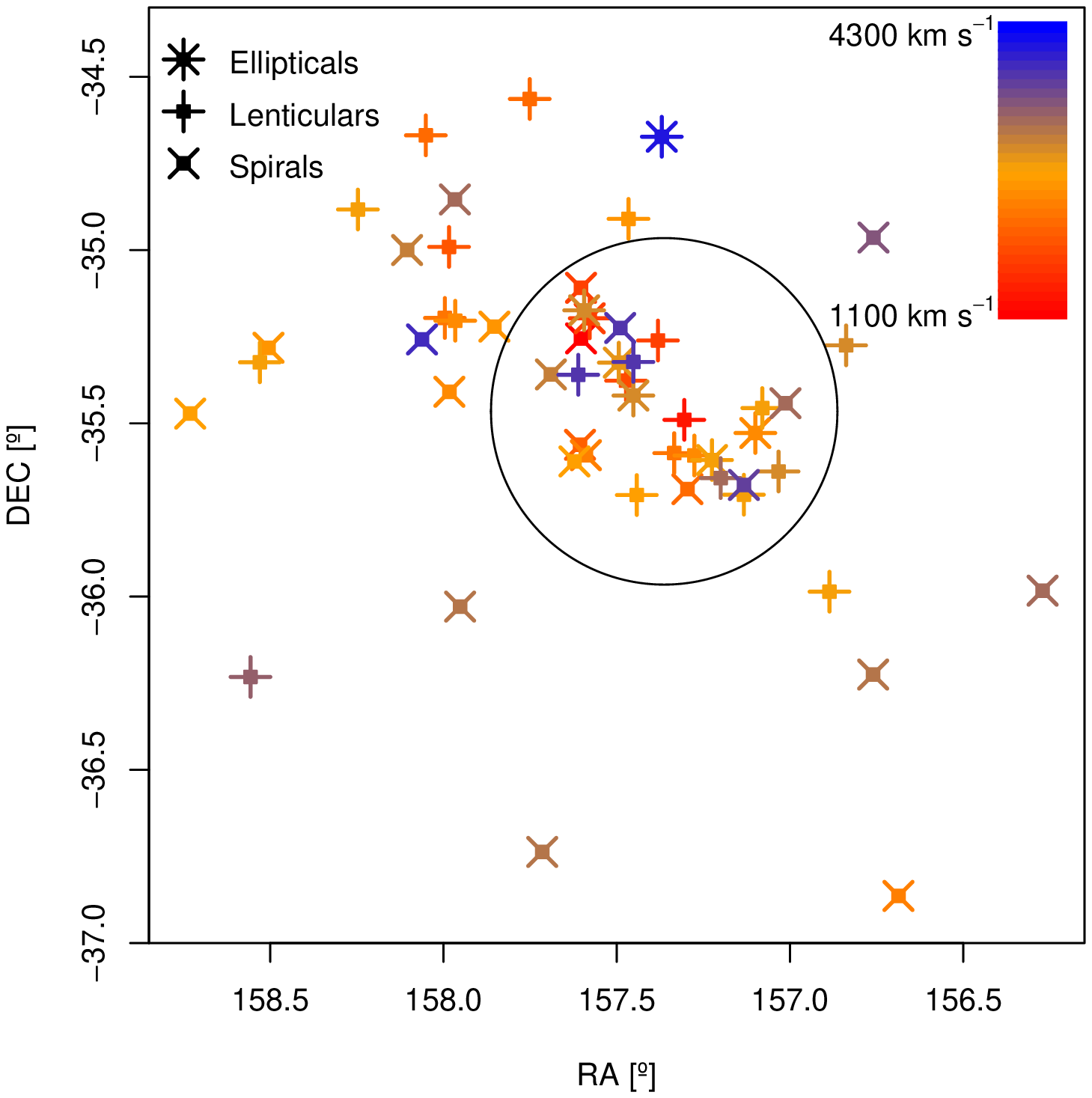}    
\includegraphics[width=60mm]{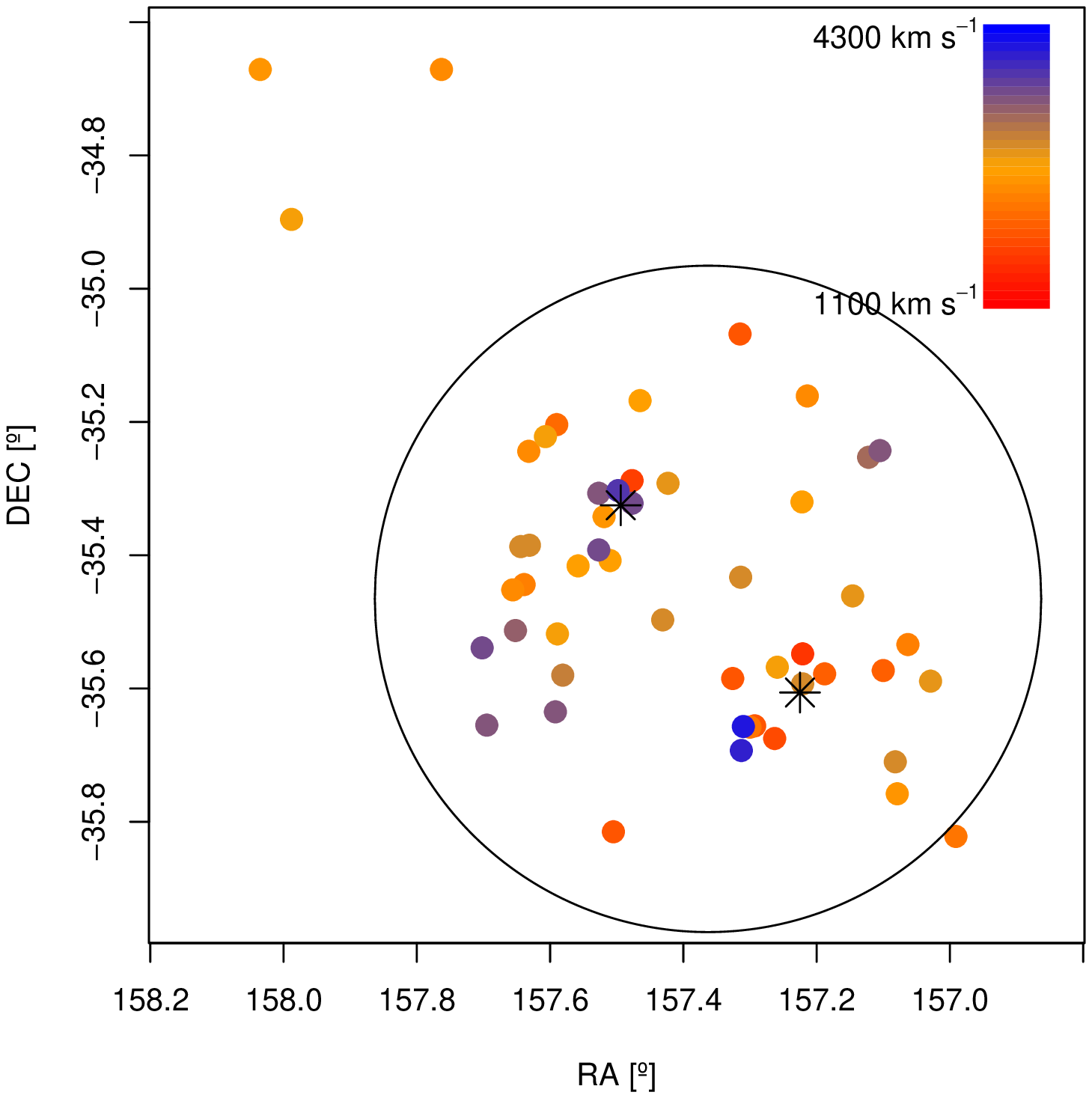}    
\caption{{\bf Left panel:} projected density distribution for 
galaxies, split into three percentile ranges. {\bf Middle panel:} projected 
positions for bright galaxies. The colour palette ranges from red to blue, 
spanning the $V_{R,h}$ range defined in Sect.\,\ref{sec:reduc}.
{\bf Right panel:} similar plot for dwarf galaxies in the sample.
The black asterisks indicate the positions of NGC\,3258 (south-west) 
and NGC\,3268 (north-east). In both panels, the origin of the black 
circle is the {\it \emph{bona fide}} centre of the cluster, and its radius 
is $30'$.}
\label{espac.typ} 
\end{figure*}    

Several studies in galaxy clusters found different velocity
dispersions for the dwarf and bright galaxies populations 
\citep[e.g.][]{dri01,con01,edw02}. To test whether this 
is also true for the Antlia cluster, we subdivided our sample in giants 
and dwarfs using the morphological classification from 
\citet{smi08a,smi12} and, when this is  not available, from \citet{fer90}.
Normal distributions were fitted by least squares to the dwarf and 
bright galaxies, separately. The resulting mean values for the bright and 
dwarf populations were $2768\pm104$\,km\,s$^{-1}$ and 
$2698\pm108$\,km\,s$^{-1}$, respectively. The corresponding dispersions 
are $698\pm122$\,km\,s$^{-1}$ and $682\pm120$\,km\,s$^{-1}$. 
From a K-S test, we cannot reject the hypothesis that both samples 
are drawn from the same distribution with $90\,\%$ confidence. 

\subsection{Morphological types and spatial distribution}
\label{morph}

Assuming that the groups of galaxies dominated by the gEs as the 
main substructures of the cluster and that both haloes present 
similar masses \citep{ped97,nak00}, we consider the midpoint
between them as its approximate centre 
($\alpha= 10^h\,29^m\,27^s$, $\delta= -35^o\,27'\,58''$).

The 105 galaxies in our sample occupy a region of $\sim 7\,{\rm degree}^2$.
Therefore, we obtained the mean surface density, $\sim15\,{\rm degree}^{-2}$, 
and the  mean distance between galaxies, $\sim9$\,arcmin. And for each 
galaxy, we calculated the number of neighbours nearer than this  value, and we 
used it to split the galaxies in four percentiles. The left-hand panel of 
Figure\,\ref{espac.typ} shows the regions occupied by the 25, 50,
and 75 percentiles of galaxies with a larger number of neighbours. The
white asteriks correspond to the position of NGC\,3258 (south-west)
and NGC\,3268 (north-east). 
The origin  of the black circle is the {\it \emph{bona fide}} centre of the 
cluster, and its radius is $30'$ (i.e. $\sim 300$\,kpc at Antlia 
distance). The 25 percentile seems to clearly represent the 
overdensities of galaxies around both gEs, which are also identified
in the 50th percentile. The lower percentile shows 
a more extended spatial distribution. 

In the middle panel of Figure\,\ref{espac.typ},  the projected 
spatial distribution for bright galaxies (i.e., all those galaxies
not classified as dwarfs) is plotted. Different symbols indicate the  Hubble type 
(i.e., spirals, lenticulars, or ellipticals). The colour palette ranges 
from red to blue, spanning the $V_{R,h}$ range defined in 
Sect.\,\ref{sec:reduc}. The origin and radius of the black circle are 
identical to those in the left-hand  panel. The two groups around each of the gEs 
can be clearly identified. There are also several galaxies located in 
the extrapolation of the straight line joining the two gEs, mainly to 
the north-east. 
These galaxies, as well as those located in the central part of the
cluster, mainly present intermediate $V_{R,h}$.  There are a few galaxies 
with $V_{R,h}<2000$\,[km\,s$^{-1}$], whose projected positions agree with 
 the general scheme of galaxies with intermediate velocities.
However, the galaxies with the higher $V_{R,h}$ seem to present a different 
projected spatial distribution. The majority of them were classified as 
spirals, despite the relatively isolated elliptical FS90-152 (see 
Table\,\ref{tab.mem}).
On the other hand, the few other ellipticals have a projected distance 
to the centre of the cluster (d$_{\rm p}$) lower than $25'$ and present 
$1775 < V_{R,h}\,[$km\,s$^{-1}] < 3000$.

The right-hand panel of Figure\,\ref{espac.typ} shows the projected spatial 
distribution for dwarf galaxies in the sample and the positions of 
NGC\,3258 and NGC\,3268. The origin and 
radius of the black circle are identical to the previous panels. Their 
concentration within the region restricted by d$_{\rm p}= 30'$ is 
due to a bias in the spectroscopic survey. Therefore, we cannot arrive 
at any conclusion regarding their distribution when d$_{\rm p}> 30'$. 
The dwarfs near the centre present a wide range of $V_{R,h}$. They seem 
to be more concentrated towards the two gEs and do not seem to be aligned 
with the projected position of the two gEs.

\begin{figure}
\includegraphics[width=90mm]{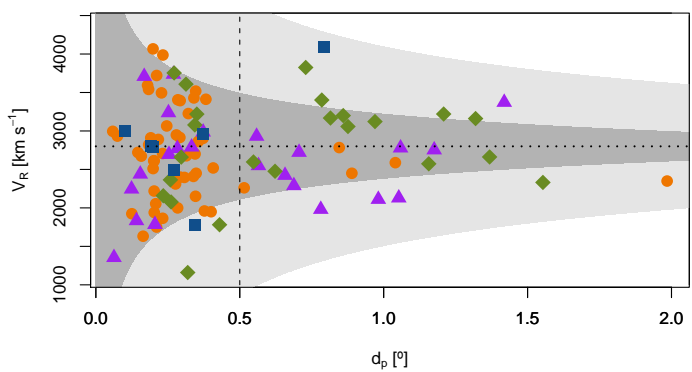}    
\caption{Heliocentric radial velocities ($V_{R,h}$) as a function of projected 
distances to the galaxy centre (d$_{\rm p}$). Different symbols represent
dwarf (red circles), ellipticals (blue squares), lenticulars (blue-violet 
triangles), and spirals (green diamonds). The dashed line indicates 
d$_{\rm p}= 30'$, and the dotted one represents the $V_{R,h}$ for both gEs, 
$\sim2800$\,km\,s$^{-1}$. The grey regions indicate the caustic curves obtained
with the \citet{pra94} infall model for two sets of parameters.}    
\label{dist_vel} 
\end{figure}

Figure\,\ref{dist_vel} shows the $V_{R,h}$ as a function of d$_{\rm p}$ 
for the galaxies in the sample, discriminated by morphological types. 
The grey regions represent caustic curves
obtained with the infall model of \citet{pra94}, assuming $\Omega_0=0.3$
and two sets of parameters.
For the inner region, we proposed typical parameters for the Fornax
cluster $r_{vir}=0.7$\,Mpc and $\sigma_{v}=374$\,km\,s$^{-1}$
\citep{dri01}. Considering the high mass derived for the Antlia cluster
by \citet{hes15}, we plotted the second region assuming the Virgo cluster
virial radius, $r_{vir}=1.8$\,Mpc \citep{kim14} and 
$\sigma_{v}=617$\,km\,s$^{-1}$, which were previously derived
in Section\,\ref{sec:dvel}. While the first set of parameters implies that
a large number of galaxies lie outside of the caustic curves, for the 
 second one all galaxies present  $V_{R,h}$-values lower than the escape velocity
at their corresponding projected distances from the assumed centre of the cluster.

Dwarf galaxies seem to be spread all over the $V_{R,h}$ range for d$_{\rm p}< 20'$. 
Just a few dwarf galaxies outside this limit were measured, but all of them 
present intermediate values of $V_{R,h}$. The picture is quite different 
when we look at the bright galaxy population. As expected, spiral galaxies 
are mainly at larger distances than $20'$ from the {\it \emph{bona fide}} cluster 
centre. Moreover, their velocity distribution also distinguishes them from
the rest of the galaxies in the sample. From the thirteen spirals with 
d$_{\rm p}> 30'$, seven of them present $V_{R,h}$ around 3200\,km\,s$^{-1}$ 
and relatively low dispersion. In fact, the mean $V_{R,h}$ and its dispersion
for these spirals are $3190\pm40$\,km\,s$^{-1}$ and 110\,km\,s$^{-1}$.

If we consider all the spirals with d$_{\rm p}> 30'$, their mean  $V_{R,h}$ 
is $2980\pm120$\,km\,s$^{-1}$. For comparison, the eleven lenticulars in the 
same radial regime have a mean velocity of $2550\pm125\,$km\,s$^{-1}$. A 
Student's test between the two samples rejects the hypothesis that both 
samples belong to the same population with  $90\,\%$ of confidence.

\begin{figure}    
\includegraphics[width=90mm]{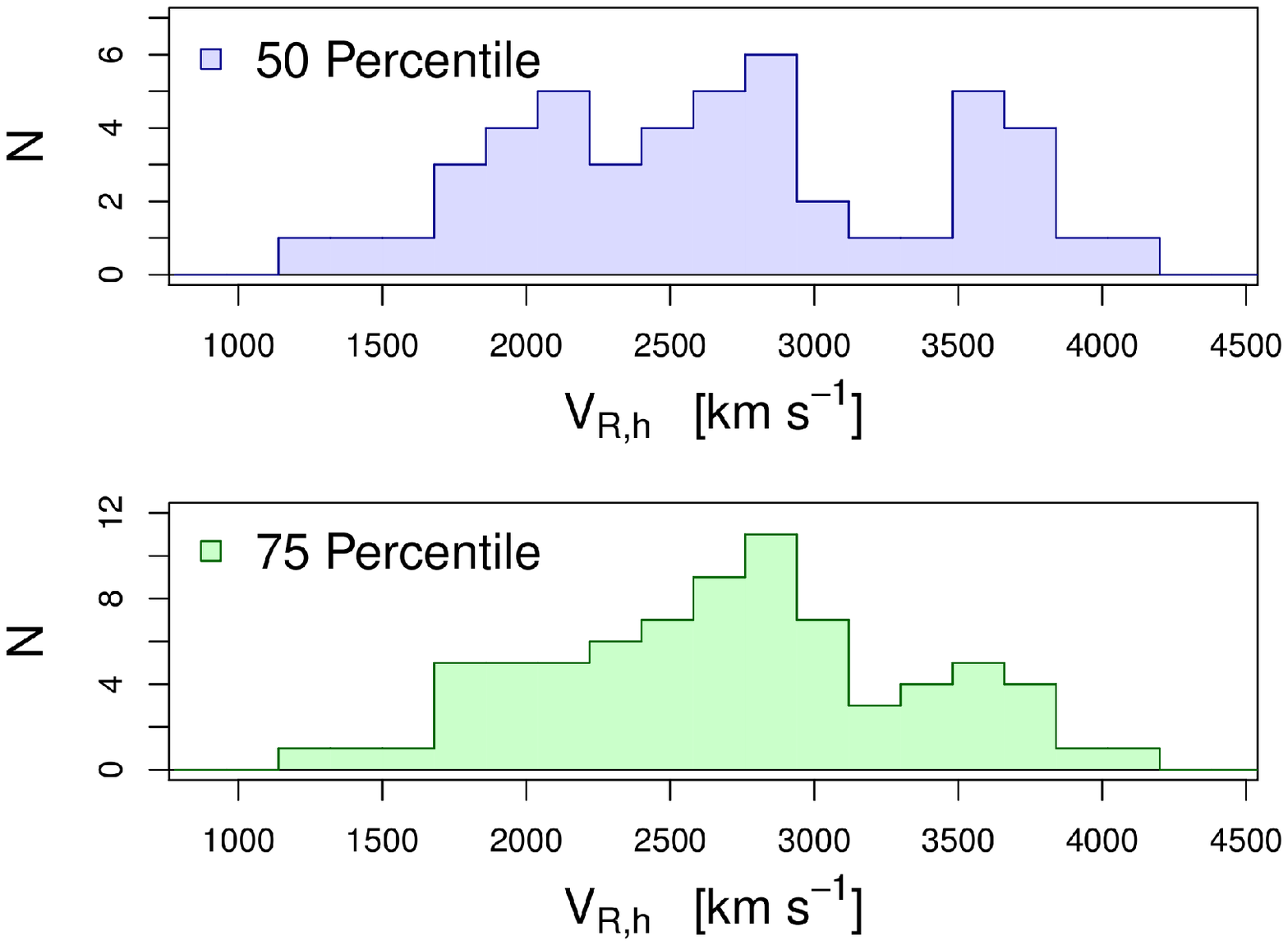}
    
\includegraphics[width=90mm]{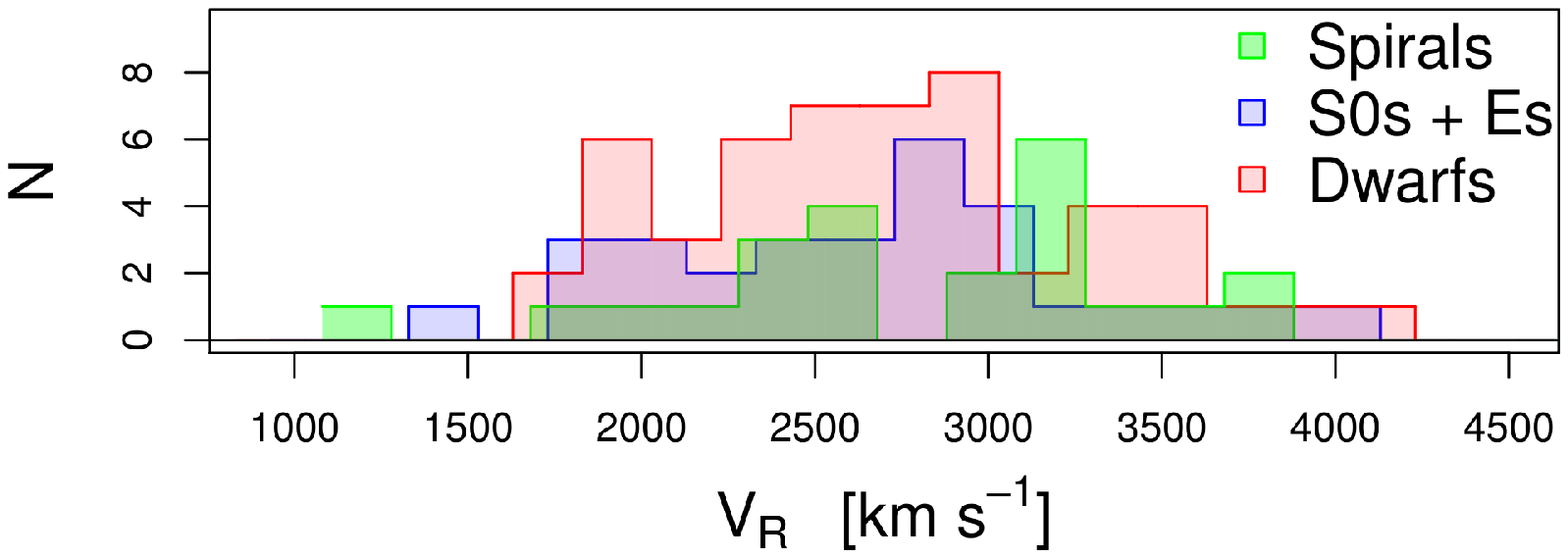}    
\caption{{\bf Upper panel:} $V_{R,h}$ distribution for galaxies
belonging to the  50 percentile with larger number of neighbours.
{\bf Middle panel:} $V_{R,h}$ distribution for galaxies
belonging to the 75 percentile with larger number of neighbours. 
{\bf Lower panel:} $V_{R,h}$ distribution for Antlia galaxies, 
separated by morphology.}    
\label{dvel2}    
\end{figure}

To study the central part of the cluster, which is expected to
be dominated by the gEs haloes, we restricted our sample to include 
the 50th percentile of galaxies with larger number of neighbours. The 
upper panel of Figure\,\ref{dvel2} shows the $V_{R,h}$ distribution
for these galaxies, adopting a bin width of 180\,km\,s$^{-1}$. It seems
to represent three groups of galaxies with $V_{R,h}$ around $\sim2000$\,km\,s$^{-1}$,
$\sim2800$\,km\,s$^{-1}$, and $\sim3700$\,km\,s$^{-1}$. To the last 
group belong the three bright lenticulars NGC\,3267 (FS90-168), NGC\,3269 
(FS90-184), and NGC\,3271 (FS90-224), near to NGC\,3268 in projected distance.
Considering the $V_{R,h}$ for both gEs, the galaxies in the group with 
intermediate velocities have the higher probability of belonging to the Antlia 
cluster. We applied a S-W test to this sample and obtained a p value of 0.32.
Then, we ran {\textsc GMM} \citep{mur10} for a trimodal case. {\textsc GMM}
calculated three peaks with means of $2020\pm160$\,km\,s$^{-1}$, $2750\pm100$\,km\,s$^{-1}$,
and $3650\pm50$\,km\,s$^{-1}$. For the unimodal hypothesis, we obtained
$\chi^2=13.9$ with eight degrees of freedom ($p\chi^2=0.1$) and a distribution kurtosis
$k=-0.9$. These results point to a multi-modal distribution. The parameter $DD$,
which measures the relevance of the peak detections, was $2.8\pm0.7$ 
(\citep[$DD>2$ is required for a meaningful detection (][]{ash94,mur10}.

In the middle panel we present the $V_{R,h}$ distribution for galaxies up
to the 75th percentile with the same bin width as in the previous plot. The 
three different groups also seem to be present, but smoothed. In this case,
the p value obtained from the S-W test is higher, $\sim 0.80$. For the
trimodal case, the mean from {\textsc GMM} were $2060\pm200$\,km\,s$^{-1}$, 
$2780\pm100$\,km\,s$^{-1}$ and $3600\pm130$\,km\,s$^{-1}$. In this case,
$\chi^2=7$ ($p\chi^2=0.5$), $k=-0.5,$ and $DD=2.5\pm0.8$. The parameters 
are less conclusive, but point to a multi-modal distribution in $V_{R,h}$.

The lower panel of Fig.\,\ref{dvel2} shows the $V_{R,h}$ distribution 
for Antlia galaxies, separated between dwarf and bright galaxies, 
and the latter group between early-types (ellipticals and lenticulars)
and late-types (spirals). Spirals fill a wide range of $V_{R,h}$, but
seem to avoid  velocities similar to those of the gEs ($\sim 2800$\,km\,s$^{-1}$).

\subsection{Mass estimations}

\citet{ped97} used X-ray observations (ASCA) to measure the mass 
enclosed within a radius of 240\,kpc from NGC\,3258 and obtained 
$0.9--2.4 \times 10^{13}\,{\rm M_{\odot}}$. NGC\,3268 was studied in 
X-rays (ASCA) by \citet{nak00}, who determined 
$\sim 2\times 10^{13}\,{\rm M_{\odot}}$ internal to $\sim 260$\,kpc.
Here we  compare our galaxy velocities with the results from X-ray 
observations. We excluded spirals because of the differences that we found
between their $V_{R,h}$ distribution and those of early-type bright 
galaxies in Section\,\ref{morph}. 
Two subsets of our sample were made, selecting from the galaxies contained
in the 50th percentile those with shorter projected distance to 
NGC\,3258 or NGC\,3268. These results in samples of 17 (up to 10\,arcmin)
and 24 members (up to 18\,arcmin), respectively.

We applied the "tracer mass estimator" (M$_{\rm Tr}$) from \citet{eva03},

\begin{equation}
 M =    \frac{C}{G N}  \sum_i V^2_{LOS,i} R_i 
,\end{equation}

\noindent where $R_i$ and $V_{LOS,i}$ are the projected distances from the
corresponding gE and velocities relative to it, respectively. Here, G is the constant of 
gravitation, and N the number of tracers. In the case of isotropy, the constant C 
is calculated through

\begin{equation}
C = \frac{ 4\,(\alpha + \gamma)}{\pi} \frac{4 - \alpha - \gamma}{3- \gamma}
\frac{1-(r_{in}/r_{out})^{3-\gamma}}{1-(r_{in}/r_{out})^{4-\alpha-\gamma}}
.\end{equation}

We assume that the three-dimensional profile of the tracer population is 
represented well by a power law between an inner radius $r_{in}$ and outer 
radius $r_{out}$ with exponents $\gamma$ and $\alpha$, respectively (e.g.
see \citealt{mam05a,mam05b}). Under 
the assumption of constant circular velocity, $\alpha=0$. In the case of the 
tracer population, it is difficult to measure the power law exponent precisely, 
but it should not differ too much from $\gamma=3$ \citep{dek05,agn14,cou14}. 
Taking this into account, we obtain the mass that corresponds to $\gamma$ ranging from 2.75 to 
3.25. In the case of a shallower profile, the resulting masses are somewhat lower,
but not by an order of magnitude.

For the sample surrounding  NGC\,3258 we obtained 
M$_{\rm Tr}= (8 - 11)\times 10^{13}\,{\rm M}_{\odot}$.
This has to be compared with the X-ray mass of NGC\,3258. \citet{ped97} 
obtained $kT_e\approx 1.7$\,keV, assuming an isothermal gas. 
The parameters of their beta model are $r_c=8.2$\,arcmin, corresponding 
to 83.5 kpc, and $\beta=0.6$.

With the assumption of spherical symmetry, we apply the expression \citep{gre01}

\begin{equation}
M(r) =  \frac{3 \beta k T_e}{G \mu m_p} \frac{r^3}{r_c^2+r^2}
\label{eq:xray}
\end{equation}

Adopting $\mu=0.6$, the derived mass within the sphere with radius 
$10'$ is $\sim 7 \times 10^{12}\,{\rm M_{\odot}}$.
In the case of the sample surrounding NGC\,3268, the result from the 
\citet{eva03} estimator is
M$_{\rm Tr}= (5 - 8)\times 10^{13}\,{\rm M}_{\odot}$.

\medskip

We again applied Eq. \ref{eq:xray} with the parameters derived by
\citet{nak00}: $kT_e= 2$\,keV, $r_c= 5$\,arcmin, and $\beta=0.38$. 
Then the mass enclosed within $18'$  around NGC\,3268 is 
$\sim 1.4 \times 10^{13}\,{\rm M_{\odot}}$.
In both cases, the mass derived from X-ray observations is significantly 
lower than the mass estimated from the $V_{R,h}$ measurements. 
It is unlikely that the mass enclosed within $\sim 100$\,kpc around 
both gEs reaches values of a few times  $10^{13}\,{\rm M_{\odot}}$. We therefore conclude that the galaxies with the most deviant velocities in the two
samples are probably not gravitationally bound to either of the two groups.

From the $V_{R,h}$ histograms in Fig\,\ref{dvel2}, a large number of 
the galaxies is symmetrically grouped around 
$2750--2800\,{\rm [km\,s^{-1}]}$, spanning approximately the 
velocity range $2200 < V_{R,h}\,{\rm [km\,s^{-1}]} < 3400$. This group
corresponds to $\sim 50\%$ of the galaxies that belong to the 50th percentile,
and $\sim 60\%$ of the galaxies for the 75th percentile.
This velocity constraint reduces the size of the samples around NGC\,3258 and
 NGC\,3268 to 8 and 13 members, respectively.
Applying the mass estimators to these limited samples, we obtain, for
the galaxies, around NGC\,3258 and NGC\,3268
M$_{\rm Tr}= (1.4 - 2)\times 10^{13}\,{\rm M}_{\odot}$ up to $9'$ and
M$_{\rm Tr}= (1.7 - 2.4)\times 10^{13}\,{\rm M}_{\odot}$ up to $15'$, 
respectively.

These estimations are in reasonable agreement with the masses derived
from X-ray observations. The mass that we derived from the analysis of an 
UCD sample around NGC3268 is $2.7\times10^{12} M_\odot$ within 47 kpc. 
Assuming a constant circular velocity, the extrapolated mass within  
101 kpc (corresponding to $10'$) is $6\times10^{12} M_\odot$, and 
 therefore also is in reasonable agreement with our mass estimation 
from both companion galaxies and X-rays. To avoid the need to explain 
peculiar radial velocities of the order of 1500\,km\,s$^{-1}$, we conclude 
that the large velocity dispersion of our complete sample is caused by 
recessional velocities of galaxies in the near foreground or background, 
which are mixed in.

\subsection{Tests for cluster substructure}
In this section, we try a different approach. 
The nearest-neighbour tests are commonly used 
for detecting subgroups in the environment of clusters of galaxies
\citep[e.g.][]{bos10,bos06,bur04,owe09,hou12}. Our aim is to search 
for substructure in $V_{R,h}$ besides the obvious existence of the two 
groups dominated by NGC\,3258 and NGC\,3268.

The $\Delta$ test \citep{dre88} was intended for probing deviations
from the local mean velocities and dispersions compared with the global
cluster values. To achieve this, the local mean velocity and dispersion
are calculated for each galaxy, restricting the sample to the galaxy
itself and its ten nearest neighbours. Then, their deviation from the 
global cluster values are computed, and their sum is defined as the 
$\Delta$ parameter.
\citet{col96} proposed the $\kappa$ test, similar in intent to the 
$\Delta$ test. The possibility that the scale in which the substructure 
is more obvious differs from what is imposed by the ten nearest neighbours 
condition is considered, leaving the number of neighbours $n$ as a free
parameter. Arguing that distributions cannot be 
characterised by the first two moments in general, they proposed a
statistic ruled by the probability of the K-S two-sample distribution
\citep{kol33,smi48}.
For both tests, the significance of their statistics are estimated by
Monte Carlo simulations, in which the velocities of the cluster galaxies
are shuffled randomly. The significance level $p$ was obtained after
repeating the previous procedure 1\,000 times.

\citet{pin96} found that the existence of radial gradients in the 
velocity dispersion of galaxies could produce artificial substructure 
when applying 3D statistics.
We therefore ran the tests on galaxies whose projected distances to the 
equidistant point to both gEs was less than $20'$, because velocity
dispersions seem to remain constant up to that limit (see 
Fig.\,\ref{dist_vel}). This restriction reduced our sample to 57 members. 
The area matches the field of view of our VIMOS survey, which 
guarantees a homogeneous sample. For the $\kappa$ test, we
selected $n=10,15$. The results were $\Delta= 66.9$ and 
$p_{\Delta}=0.12$, $\kappa_{10}=14.4$ and $p_{10}=0.16$, $\kappa_{15}=15.6$ 
and $p_{15}=0.06$.

To investigate the significance of these numbers, we performed 1\,000 Monte 
Carlo simulations. We randomly generated the $V_{R,h}$ for the galaxies 
in the sample, assuming a normal distribution with mean and dispersion of 
2800\,km\,s$^{-1}$ and 275\,km\,s$^{-1}$, respectively.
Then, we applied the tests to these samples. For the $\Delta$ test,
fewer than $15\%$ of the cases produced a $\Delta$-parameter higher
than the value obtained from the observations, and this percentage is 
reduced to $6\%$ for the $\kappa_{15}$ test.
 Therefore, the tests point to underlying substructure, most probably due 
to the existence of non-member galaxies in our sample.

\section{Discussion}

\subsection{Extreme radial velocities}

To use radial velocities as a discriminant for cluster membership is often difficult 
if the depth along the line of sight is considerable. An extreme example is the interesting 
pair of Virgo galaxies IC\,3492 with a radial velocity of -575 km/s and IC\,3486 with 1903 km/s, 
having  a separation of only 1.4 arc min. The deviation from a strict Hubble law may be a 
mix of peculiar velocities resulting from large scale structure and local gravitational fields. 
Is it possible that the extreme radial velocities in Fig.\ref{dist_vel} are infall velocities?
To answer this, it is useful to consider  the highest possible velocities. We  can 
represent the total mass distribution around NGC\,3268 quite well by an NFW-mass distribution with 
a scale length of 22 kpc and a characteristic density of 0.05$M_\odot/pc^3$. A mass probe on an 
exactly radial orbit, initially at rest and falling in from a distance of 1.5 Mpc, crosses the 
centre after 6.8 Gyr with a velocity of 1550\,km\,s$^{-1}$. However, after an additional 0.06 Gyr, its
velocity has already declined to 1000\,km\,s$^{-1}$. It is obvious that such a configuration is 
highly artificial and not suitable to explaining the broad velocity interval.
The more plausible explanation is therefore the mix of recessional and Doppler velocities.

 \subsection{Internal structure of the two dominant groups}
 
What we called the Antlia cluster seems to be mainly formed by two subgroups, each one 
dominated by a gE. Despite that, both galaxies present similar $V_{R,h}$, it is uncertain 
whether they are located at the same distance \citep[e.g.][]{bla01,can05,bas08}. 
A recent study in HI \citep{hes15} indicates that the subgroup dominated by NGC\,3258 
could be falling in to NGC\,3268, which they estimated is the actual cluster centre.
It is clear that both gEs dominate their own subgroups, and their peculiar velocities
relative to the group systemic velocity should not differ significantly. For this reason, if
the difference in distance between both galaxies was $2-4\,Mpc$, we would expect a
$V_{R,h}$ difference of $\sim 140-280$\,km\,s$^{-1}$. As a result, the
measured $V_{R,h}$ might agree with the scenario suggested by \citet{hes15},
where both subgroups are in a merging process.

It is  interesting to have a closer look at the group around NGC\,3268. Strikingly, all 
three S0's around the central galaxy (FS90-224, FS90-168, and FS90-184) have positive 
velocity offsets in the range of $900-1000$\,km\,s$^{-1}$. If these velocities were due 
to the potential of NGC\,3268 (and its associated dark matter), very radial orbits would be
needed for surpassing the circular velocity that is about 300\,km\,s$^{-1}$ and probably 
constant \citep{cas14}. The high velocity offsets are, moreover, related to fairly large 
projected spatial offsets (which in the case of FS90-224 is about 65\,kpc), whereas the 
highest velocities are found at the bottom of the potential well. Furthermore, the 
galaxies do not show any morphological indication of being tidally disturbed. (We comment 
separately on the special case NGC\,3269.)  Moreover, there is no indication that the 
globular cluster system of NGC\,3268 is tidally disturbed \citep{dir03b}, nor does the 
X-ray structure show any abnormality \citep{nak00}. We therefore prefer the interpretation 
that the three S0s have to be placed  in the background of NGC\,3268. As Fig.\ref{dvel2} 
suggests, more dwarf galaxies can be associated with this group, but some cases are 
ambiguous. There is the dwarf FS90-195 that hosts about ten point sources, which may be 
globular clusters. 
The radial velocity is 3495\,km\,s$^{-1}$, but it is located close to NGC\,3268 (projected 
distance 15\,kpc), so it may be bound and demonstrates how globular clusters can be donated 
to the system of NGC3268. In that sense it could resemble SH2 in NGC1316 \citep{ric12c}. 

We note that NGC\,3269 (FS90-184), together with NGC\,3267 (FS90-168)  and NGC\,3271 
(FS90-224), has been identified as the "Lyon Group of Galaxies (LGG)" 202 of 
\citet{gar93,gar95} and \citet{bar01}, who already suspected its location behind the Antlia 
cluster.

Galaxies of the low velocity group are mostly dwarf-like galaxies and might be {\it \emph{bona fide}}
in the foreground. All look quite discy, which may also indicate a less dense environment.
The situation is somewhat different for the group around NGC\,3258. There are also dwarfs 
(ANTLJ102914-353923.6 and FS90-137) with positive velocity offsets of more than 
1000\,km\,s$^{-1}$ to NGC\,3258, but the brighter galaxies do not show such extreme radial 
velocities. There is also a group around FS90-226 that is less striking than the two dominant 
groups, which is also identified as a subgroup by \citet{hes15}.

Therefore the likely explanation for Fig.\ref{dist_vel} is that we are looking along a 
filament of galaxies and that at least a fraction of the galaxies in the three velocity 
groups apparent in Fig.\ref{dist_vel} are separated by their recession velocities. 

\subsection{NGC 3268/3258 as fossil groups?}

If the crowding of galaxies around NGC 3268 is mainly a projection effect, this galaxy 
fulfils the criteria for being a fossil group. After applying the definition of \cite{jon03}, 
it has the required X-ray luminosity of more than $10^{42}$\,erg\,s$^{-1}$, and the next 
galaxy in a luminosity ranking would be FS90-177, which is fainter than 2\,mag in 
the $R$ band.   

The brighter galaxies  around NGC\,3258 (FS90-105 and FS90-125) may be companions. 
FS90-125 is perhaps a magnitude fainter than NGC\,3258, so only the X-ray criterium applies, 
and it cannot be called a fossil group. However, its globular cluster system is even 
richer than that of NGC\,3268, so the idea that a galaxy group collapsed at very 
early times is plausible. 

\subsection{Emission line galaxies}
Only three galaxies from our VIMOS sample present emission lines. One of 
them, FS90-131, was classified as a spiral. The other galaxies, 
FS90-220 and FS90-222, were classified as lenticulars. Evidence of
strong star formation would not be expected  if the environment 
 of the Antlia gEs was dense: FS90-131 is located at $5'$ from NGC\,3258, 
FS90-220 is $\sim 6'$ from NGC\,3268, and FS90-222 is 
$\sim 3'$ from NGC\,3273. Moreover, the $V_{R,h}$ in the three
cases is lower than 1200\,km\,s$^{-1}$, which disqualifies them from dynamically
belonging to these groups. Their velocities are better explained by recession
velocities.

\section{Summary and conclusions}
We have presented new radial velocities, measured with VLT-VIMOS and Gemini GMOS-S, 
for galaxies in the region normally addressed as the Antlia cluster. The fields are 
located in the surroundings of NGC\,3258 and NGC\,3268, the dominant galaxies of 
the cluster. Together with the literature data, our list of Antlia galaxies with 
measured radial velocities now embraces 105 galaxies. Large numbers of these objects 
are projected onto either of the two subgroups around NGC\,3258 or NGC\,3268. Because 
the gravitational potentials of these galaxies are constrained by X-ray studies and 
stellar dynamical tracers, we could compare the observed radial velocities with those 
expected. It turned out that the total range of velocities seems to be too high to 
be generated by the gravitational potential of NGC\,3258 or NGC\,3268. There are three 
groups of galaxies (not always clearly separated) characterised by radial velocities 
of about 1800\,km\,s$^{-1}$, 2700\,km\,s$^{-1}$, and 3700\,km\,s$^{-1}$. We interpreted 
these velocities as recession velocities, which places the bright S0s around NGC\,3268, 
in particular, into the background. Therefore, NGC\,3268 might qualify as a fossil group, 
resembling, for example, NGC\,4636 in the Virgo cluster in its properties. The intermediate 
velocity group (i.e.$V_{R,h}$ around 2700\,km\,s$^{-1}$) is the most populated one. The 
distances between NGC\,3258 and NGC\,3268 are not well constrained in the literature, and 
we cannot discard that the groups dominated by them are in the early stages of a merging 
process.

What we originally called the Antlia cluster is therefore characterised by several 
groups that, at least in some cases, are not gravitationally bound. We may be 
looking along a filament of the cosmic web.

\begin{acknowledgements}
We thank Ylva Schuberth for discussions about the radial velocities of Virgo 
galaxies, Mike Fellhauer for permission to use his orbit program, Francisco 
Azpillicueta for discussions of statistical issues, and Lilia Bassino for discussions
about the Antlia cluster. We thank the refereee for suggestions that improved this article.\\
This work was based on observations made with ESO telescopes at the La Silla Paranal Observatory under 
programmes ID 60.A-905 and ID 079.B-0480, and on observations obtained at 
the Gemini Observatory, which is operated by the Association of Universities 
for Research in Astronomy, Inc., under a cooperative agreement with the NSF on 
behalf of the Gemini partnership: the National Science Foundation (United 
States), the National Research Council (Canada), CONICYT (Chile), the 
Australian Research Council (Australia), Minist\'{e}rio da Ci\^{e}ncia, 
Tecnologia e Inova\c{c}\~{a}o (Brazil) and Ministerio de Ciencia, 
Tecnolog\'{i}a e Innovaci\'{o}n Productiva (Argentina), and on observations
acquired through the Gemini Science Archive. This research has made use of the NASA/IPAC 
Extragalactic Database (NED), which 
is operated by the Jet Propulsion Laboratory, California Institute of Technology, 
under contract with the National Aeronautics and Space Administration.\\
This work was funded with grants from Consejo Nacional de Investigaciones   
Cient\'{\i}ficas y T\'ecnicas de la Rep\'ublica Argentina, Agencia Nacional de   
Promoci\'on Cient\'{\i}fica y Tecnol\'ogica (BID AR PICT-2013-0317), and Universidad Nacional de La Plata  
(Argentina). TR is grateful for financial support from FONDECYT project Nr.\,1100620, 
and from the BASAL Centro de Astrof\'isica y Tecnolog\'ias Afines (CATA) PFB-06/2007.
\end{acknowledgements}

\begin{table*}[ht]
\begin{center}   
\label{tab.backg}    
\caption{Heliocentric radial velocities for background galaxies measured in 
this paper.}
\begin{tabular}{lccc}
\hline    
\\    
\multicolumn{1}{l}{ID}&\multicolumn{1}{c}{RA(J2000)}&\multicolumn{1}{c}{DEC(J2000)}&\multicolumn{1}{c}{$V_{\rm R,h}$}\\    
\multicolumn{1}{l}{}&\multicolumn{1}{c}{hh mm ss}&\multicolumn{1}{c}{dd mm ss}&\multicolumn{1}{c}{km\,s$^{-1}$}\\    
\hline    
\\   
FS90-75&10 28 12.0&-35 32 20.4&12333$\pm$07\\
FS90-205&10 30 18.5&-35 24 43.2&45907$\pm$30\\
ANTL102823-352754&10 28 23.8&-35 27 54.0&40289$\pm$25\\
ANTL10294-352320&10 29 04.6&-35 23 20.4&51147$\pm$18\\
ANTL10295-352146&10 29 05.5&-35 21 46.8&52928$\pm$23\\
ANTL102817-353552&10 28 17.0&-35 35 52.8&44240$\pm$85\\
ANTL102937-353552&10 29 37.2&-35 35 52.8&7682$\pm$55\\
ANTL102936-353336&10 29 36.7&-35 33 36.0&24420$\pm$50\\
ANTL102951-351210&10 29 51.6&-35 12 10.8&32797$\pm$24\\
ANTL10303-3571&10 30 03.8&-35 7 01.2&28192$\pm$21\\
ANTL103042-35914&10 30 42.2&-35 9 14.4&15875$\pm$14\\
ANTL103049-352031&10 30 49.4&-35 20 31.2&32830$\pm$23\\
ANTL103045-35161&10 30 45.8&-35 16 01.2&32920$\pm$21\\
ANTL102945-35374&10 29 45.4&-35 37 04.8&8900$\pm$19\\
ANTL102936-353336&10 29 36.7&-35 33 36.0&24337$\pm$33\\
ANTL102926-351051&10 29 26.9&-35 10 51.6&20041$\pm$10\\
ANTL103041-351250&10 30 41.0&-35 12 50.4&9833$\pm$65\\
ANTL102940-35258&10 29 40.6&-35 25 08.4&53700$\pm$42\\
ANTL102936-35266&10 29 36.0&-35 26 06.0&24243$\pm$10\\
FS90-83&10 28 23.0&-35 30 57.6&19670$\pm$18\\
FS90-88&10 28 28.1&-35 31 04.8&19624$\pm$74\\
ANTL102838-352027&10 28 38.2&-35 20 27.6&46732$\pm$54\\
\hline    
\end{tabular}    
\end{center}    
\end{table*}

\bibliographystyle{aa}
\bibliography{biblio}

\onecolumn
\begin{center}
\begin{longtable}{cccccc}
\caption{Heliocentric radial velocity for Antlia members up to date.} 
\label{tab.mem} \\

\hline
\\
\multicolumn{1}{c}{ID} & \multicolumn{1}{c}{NGC} & \multicolumn{1}{c}{RA(J2000)} & \multicolumn{1}{c}{DEC(J2000)} & \multicolumn{1}{c}{Hubble Type} & \multicolumn{1}{c}{$V_{\rm R,h}$}\\ 
\multicolumn{1}{c}{} & \multicolumn{1}{c}{} & \multicolumn{1}{c}{hh mm ss} & \multicolumn{1}{c}{dd mm ss} & \multicolumn{1}{c}{} & \multicolumn{1}{c}{km\,s$^{-1}$}\\
\hline
\\ 
\endfirsthead
\caption{continued.}\\
\hline
\\
\multicolumn{1}{c}{ID} & \multicolumn{1}{c}{NGC} & \multicolumn{1}{c}{RA(J2000)} & \multicolumn{1}{c}{DEC(J2000)} & \multicolumn{1}{c}{Hubble Type} & \multicolumn{1}{c}{$V_{\rm R,h}$}\\ 
\multicolumn{1}{c}{} & \multicolumn{1}{c}{} & \multicolumn{1}{c}{hh mm ss} & \multicolumn{1}{c}{dd mm ss} & \multicolumn{1}{c}{} & \multicolumn{1}{c}{km\,s$^{-1}$}\\
\hline
\\ 
\endhead
\hline    
\endfoot

\hline \hline
\multicolumn{6}{l}{The upper 
index in column 6 indicates the reference for the $V_{R,h}$ measurement:} \\
\multicolumn{6}{l}{$^a$this paper, $^b$\citet{smi08a,smi12}}\\
\multicolumn{6}{l}{The rest  of the measurements were obtained from NED, and their references} \\
\multicolumn{6}{l}{are: $^1$ \citet{mat95}, $^2$ \citet{the98}, $^3$ \citet{jon09}} \\
\multicolumn{6}{l}{$^4$ \citet{huc12}, $^5$ \citet{dev91}, $^6$ \citet{oga08},} \\
\multicolumn{6}{l}{$^7$\citet{mat92}, $^8$\citet{lau89}, $^9$\citet{lon98}}\\
\endlastfoot

FS90-01&&10 25 05.04&-35 58 58.8&SmV&3220$\pm$05$^1$\\
FS90-18&&10 26 44.88&-36 51 50.4&SmIV&2331$\pm$05$^1$\\
FS90-28&&10 27 02.16&-34 57 50.4&SbII&3402$\pm$08$^2$\\
FS90-29&&10 27 02.40&-36 13 30.0&Sc&3122$\pm$06$^2$\\
FS90-44&&10 27 21.12&-35 16 30.0&S0&2931$\pm$45$^3$\\
FS90-50&&10 27 32.64&-35 59 09.6&S0&2722$\pm$31$^3$\\
FS90-64&&10 27 57.84&-35 49 19.2&dSB0&2261$\pm$45$^3$\\
FS90-68&&10 28 03.12&-35 26 31.2&SBab&3219$\pm$18$^a$\\
&&&&&3188$\pm$45$^3$\\
FS90-70&&10 28 06.96&-35 35 20.4&dE&2864$\pm$70$^b$\\
FS90-72&&10 28 07.92&-35 38 20.4&S0&2986$\pm$38$^b$\\
FS90-77&&10 28 15.12&-35 32 02.4&dE,N&2396$\pm$17$^a$\\
&&&&&2382$\pm$49$^b$\\
FS90-79&&10 28 19.2&-35 27 21.6&S0&2772$\pm$15$^a$\\
&&&&&2930$\pm$60$^b$\\
FS90-80&&10 28 18.96&-35 45 28.8&dS0&2519$\pm$31$^3$\\
ANTLJ102820-354236&&10 28 19.68&-35 42 36.0&dE,N&2904$\pm$40$^a$\\
FS90-84&&10 28 24.00&-35 31 40.8&E&2489$\pm$26$^a$\\
&&&&&2428$\pm$30$^4$\\
FS90-85&&10 28 24.0&-35 34 22.8&dE&2000$\pm$200$^b$\\
FS90-87&&10 28 25.2&-35 14 34.8&dE,N&3429$\pm$13$^a$\\
ANTLJ102829-351510.8&&10 28 29.3&-35 15 10.8&dE,N&3226$\pm$40$^a$\\
FS90-93&&10 28 31.92&-35 40 40.8&SmV&3608$\pm$57$^a$\\
FS90-94&&10 28 31.92&-35 42 21.6&S0&2791$\pm$24$^a$\\
&&&&&2786$\pm$45$^3$\\
FS90-98&&10 28 35.04&-35 27 39.6&BCD&2890$\pm$94$^b$\\
FS90-103&&10 28 45.12&-35 34 40.8&dE&2054$\pm$29$^b$\\
FS90-105&3257&10 28 48.0&-35 39 28.8&SB01&3237$\pm$15$^a$\\
&&&&&3200$\pm$26$^b$\\
FS90-106&&10 28 51.36&-35 09 39.6&BCD&2409$\pm$115$^b$\\
FS90-108&&10 28 53.28&-35 19 12.0&dE,N&2611$\pm$39$^b$\\
FS90-109&&10 28 53.04&-35 32 52.8&dE&1632$\pm$32$^a$\\
&&&&&1618$\pm$24$^b$\\
FS90-110&&10 28 53.04&-35 35 34.8&M32&2911$\pm$7$^b$\\
FS90-111&3258&10 28 54.00&-35 36 21.6&E&2792$\pm$50$^a$\\
&&&&&2792$\pm$28$^b$\\
FS90-120&&10 29 02.16&-35 34 04.8&ImV&2721$\pm$30$^a$\\
&&&&&2634$\pm$13$^b$\\
FS90-123&&10 29 03.12&-35 40 30.0&dE,N&1865$\pm$25$^b$\\
FS90-125&3260&10 29 06.24&-35 35 34.8&S02&2439$\pm$46$^b$\\
ANTLJ102910-353920.1&&10 29 10.32&-35 39 21.6&dE,N&1940$\pm$155$^b$\\
FS90-131&&10 29 11.04&-35 41 24.0&Sb(r)&2158$\pm$10$^a$\\
&&&&&2104$\pm$60$^b$\\
FS90-133&&10 29 12.00&-35 39 28.8&dE,N&2219$\pm$15$^a$\\
&&&&&2205$\pm$24$^b$\\
FS90-134&&10 29 13.20&-35 29 24.0&S0&1355$\pm$60$^b$\\
ANTLJ102914-353923.6&&10 29 14.40&-35 39 25.2&dE,N&4067$\pm$115$^b$\\
FS90-136&&10 29 15.36&-35 25 58.8&dE,N&2995$\pm$16$^a$\\
&&&&&2989$\pm$10$^b$\\
FS90-137&&10 29 15.12&-35 41 34.8&ImV&3987$\pm$36$^b$\\
FS90-139&&10 29 15.60&-35 04 04.8&dE,N&1950$\pm$55$^3$\\
FS90-140&&10 29 18.24&-35 35 6.0&dE,N&1924$\pm$31$^b$\\
FS90-142&&10 29 20.16&-35 35 09.6&dS0&2245$\pm$13$^b$\\
FS90-152&&10 29 28.86&-34 40 22.8&E&4093$\pm$45$^3$\\
FS90-153&&10 29 31.44&-35 15 39.6&S0&1785$\pm$13$^a$\\
&&&&&1733$\pm$39$^b$\\
FS90-159&&10 29 41.52&-35 17 31.2&dE,N&2821$\pm$21$^a$\\
FS90-162&&10 29 43.44&-35 29 49.2&dE,N&2933$\pm$31$^a$\\
FS90-165&&10 29 46.08&-35 42 25.2&S0&2695$\pm$14$^a$\\
&&&&&2604$\pm$45$^3$\\
FS90-168&3267&10 29 48.48&-35 19 22.8&SB01/2&3709$\pm$33$^b$\\
FS90-169&&10 29 48.48&-35 25 12.0&E&2999$\pm$37$^b$\\
FS90-172&&10 29 51.84&-34 54 36.0&S0&2549$\pm$19$^5$\\
FS90-173&&10 29 51.60&-35 10 04.8&dE&2677$\pm$32$^a$\\
&&&&&2609$\pm$45$^3$\\
FS90-175&&10 29 53.52&-35 22 37.2&dSB01&1834$\pm$25$^a$\\
&&&&&1781$\pm$66$^b$\\
FS90-176&&10 29 54.48&-35 17 16.8&dE,N&1751$\pm$34$^a$\\
FS90-177&&10 29 54.48&-35 19 19.2&dE,N&3540$\pm$9$^a$\\
&&&&&3505$\pm$45$^b$\\
FS90-184&3269&10 29 57.60&-35 13 30.0&S0/a&3754$\pm$33$^b$\\
FS90-185&3268&10 29 58.56&-35 19 30.0&E&2800$\pm$21$^b$\\
FS90-186&&10 29 59.52&-35 18 10.8&dE&3721$\pm$45$^a$\\
FS90-187&&10 30 01.20&-35 48 54.0&dS0&1960$\pm$55$^3$\\
FS90-188&&10 30 02.40&-35 24 28.8&dE&2673$\pm$17$^b$\\
FS90-192&&10 30 04.56&-35 20 31.2&M32&2511$\pm$18$^a$\\
&&&&&2526$\pm$4$^b$\\
FS90-195&&10 30 06.48&-35 18 25.2&dE&3495$\pm$64$^a$\\
FS90-196&&10 30 06.48&-35 23 31.2&dE&3593$\pm$9$^b$\\
ANTLJ103013-352458.3&&10 30 13.92&-35 24 57.6&dE,N&2613$\pm$200$^b$\\
FS90-208&&10 30 18.72&-35 11 49.2&S0&1774$\pm$100$^b$\\
FS90-209&&10 30 19.44&-35 34 48.0&dE&3065$\pm$13$^b$\\
FS90-212&&10 30 21.36&-35 35 31.2&SmIII&2364$\pm$27$^b$\\
ANTLJ103021-35314.8&&10 30 21.38&-35 31 04.8&dE,N&2707$\pm$19$^a$\\
FS90-213&&10 30 21.60&-35 12 14.4&dE&2151$\pm$60$^a$\\
&&&&&2185$\pm$21$^b$\\
ANTLJ103022-353806&&10 30 22.08&-35 38 06.0&dE,N&3405$\pm$40$^a$\\
FS90-216&&10 30 22.56&-35 10 26.4&E&2957$\pm$22$^a$\\
&&&&&2944$\pm$103$^b$\\
FS90-219&&10 30 24.72&-35 06 32.4&Sb&1781$\pm$45$^3$\\
FS90-220&&10 30 24.72&-35 15 18.0&S0/a&1160$\pm$8$^a$\\
&&&&&1182$\pm$45$^3$\\
FS90-222&&10 30 25.44&-35 33 43.2&S0/a&2077$\pm$6$^a$\\
&&&&&2140$\pm$45$^3$\\
FS90-223&&10 30 25.68&-35 13 19.2&dE,N&2701$\pm$29$^a$\\
&&&&&2661$\pm$9$^b$\\
FS90-224&3271&10 30 26.64&-35 21 36.0&Sb02&3737$\pm$27$^b$\\
FS90-226&3273&10 30 29.28&-35 36 36.0&S0/a&2660$\pm$08$^a$\\
&&&&&2503$\pm$20$^6$\\
FS90-227&&10 30 31.44&-35 23 06.0&dE&2948$\pm$34$^a$\\
&&&&&2921$\pm$60$^b$\\
FS90-228&&10 30 31.68&-35 14 38.4&dE,N&2450$\pm$18$^a$\\
&&&&&2417$\pm$13$^b$\\
ANTLJ103033-352638.6&&10 30 33.36&-35 26 38.4&dE,N&2311$\pm$130$^b$\\
FS90-231&&10 30 34.56&-35 23 13.2&dE,N&2915$\pm$20$^a$\\
&&&&&2909$\pm$38$^b$\\
ANTLJ103036-353046.8&&10 30 36.48&-35 30 46.8&dE,N&3394$\pm$27$^a$\\
ANTLJ103037-352708.8&&10 30 37.44&-35 27 07.2&dE,N&2400$\pm$100$^b$\\
FS90-238&&10 30 45.6&-35 21 32.4&Sm&3078$\pm$37$^a$\\
ANTLJ103047-353918&&10 30 46.8&-35 39 18.0&dE,N&3412$\pm$32$^a$\\
FS90-241&&10 30 48.48&-35 32 20.4&dE,N&3518$\pm$36$^a$\\
FS90-244&&10 30 51.60&-36 44 13.2&SBb(r)II&3161$\pm$45$^3$\\
FS90-253&&10 31 00.24&-34 33 50.4&SB0&2111$\pm$18$^5$\\
FS90-258&&10 31 03.12&-34 40 15.6&dE,N&2450$\pm$55$^3$\\
FS90-277&&10 31 24.72&-35 13 15.6&SBb(rs)II&2597$\pm$19$^5$\\
FS90-298&&10 31 48.48&-36 01 44.4&Sd&3168$\pm$38$^4$\\
FS90-300&&10 31 52.08&-34 51 14.4&Sa&3200$\pm$22$^2$\\
FS90-301&&10 31 51.84&-35 12 14.4&S0&2423$\pm$45$^3$\\
FS90-304&&10 31 55.92&-35 24 32.4&Sa&2476$\pm$16$^2$\\
FS90-306&&10 31 56.16&-34 59 27.6&SB0&1981$\pm$45$^3$\\
FS90-307&&10 31 57.12&-34 53 45.6&dE&2784$\pm$30$^a$\\
FS90-309&&10 31 59.04&-35 11 45.6&SB0&2289$\pm$19$^5$\\
FS90-318&&10 32 08.40&-34 40 15.6&dE,N&2588$\pm$45$^3$\\
FS90-321&&10 32 12.24&-34 40 08.4&SB0&2129$\pm$45$^3$\\
FS90-323&&10 32 14.88&-35 15 28.8&Sm&3825$\pm$45$^3$\\
FS90-325&&10 32 25.20&-35 00 00.0&Sd&3058$\pm$10$^7$\\
FS90-331&&10 32 59.28&-34 52 58.8&S0&2779$\pm$19$^5$\\
FS90-341&&10 34 01.20&-35 16 55.2&Sa pec&2573$\pm$08$^8$\\
FS90-343&&10 34 07.20&-35 19 26.4&S0&2754$\pm$09$^9$\\
FS90-345&&10 34 13.68&-36 13 55.2&S0&3372$\pm$45$^3$\\
FS90-353&&10 34 55.44&-35 28 19.2&Sd&2662$\pm$45$^3$\\
FS90-373&&10 37 22.80&-35 21 36.0&dE,N&2349$\pm$55$^3$\\
\end{longtable}
\end{center}
\end{document}